\documentclass{rspublic}
\usepackage{graphicx}
\usepackage{epstopdf}

\newcommand{\ea}{{\it et al. }}
\newcommand{\apj}{{\it Astrophys. J.}}
\newcommand{\apjs}{{\it Astrophys. J. Suppl. Ser.}}
\newcommand{\aj}{{\it Astron. J.}}
\newcommand{\mnras}{{\it Mon. Not. R. Astron. Soc.}}
\newcommand{\aanda}{{\it Astron. Astrophys.}}
\newcommand{\pasp}{{\it Publ. Astron. Soc. Pac.}}
\newcommand{\ptra}{{\it Phil. Trans. R. Soc. A}}

\begin{document}

\title[Young and intermediate-age massive star clusters]{Young and intermediate-age massive star clusters} 
\author[S.~S.~Larsen]{S{\o}ren S. Larsen}
\affiliation{Astronomical Institute, University of Utrecht,
Princetonplein 5, 3584 CC Utrecht, The Netherlands}
\label{firstpage}
\maketitle

\begin{abstract}{\bf galaxies: star clusters; globular clusters:
general} An overview of our current understanding of the formation and
evolution of star clusters is given, with main emphasis on high-mass
clusters. Clusters form deeply embedded within dense clouds of
molecular gas. Left-over gas is cleared within a few million years
and, depending on the efficiency of star formation, the clusters may
disperse almost immediately or remain gravitationally bound. Current
evidence suggests that a few percent of star formation occurs in
clusters that remain bound, although it is not yet clear if this
fraction is truly universal. Internal two-body relaxation and external
shocks will lead to further, gradual dissolution on timescales of up
to a few hundred million years for low-mass open clusters in the Milky
Way, while the most massive clusters ($>10^5$ M$_\odot$) have
lifetimes comparable to or exceeding the age of the Universe. The
low-mass end of the initial cluster mass function is well approximated
by a power-law distribution, $\rd N/\rd M \propto M^{-2}$, but there is
mounting evidence that quiescent spiral discs form relatively few
clusters with masses $M>2\times10^5$ M$_\odot$. In starburst galaxies
and old globular cluster systems, this limit appears to be higher, at
least several $\times 10^6$ M$_\odot$. The difference is likely
related to the higher gas densities and pressures in starburst
galaxies, which allow denser, more massive giant molecular clouds to
form. Low-mass clusters may thus trace star formation quite
universally, while the more long-lived, massive clusters appear to
form preferentially in the context of violent star formation.
\end{abstract}

\section{Introduction}

The appeal of star clusters as tools to study (extra)galactic
star-formation histories is at least two-fold: the most massive
clusters tend to be long-lived, and therefore potentially carry
information about the entire star-formation histories of their host
galaxies. Furthermore, they are bright and can be observed at much
greater distances than individual stars. Young, massive, compact
clusters have now been observed in many external galaxies, and it is
commonly assumed that at least some of these are young counterparts of
the ancient \emph{globular} clusters (GCs) which are ubiquitous in all
major galaxies. Thus, the view that GC formation required unique
physical conditions in the early Universe (e.g., Peebles \& Dicke
1968) has largely been abandoned. However, it remains much less clear
how \emph{direct} the link is between star formation and the formation
of (massive) clusters.
 
In general terms, we might ask the question, \emph{What are the
conditions for the formation and survival of massive clusters?}
Clearly, finding an answer is essential if we wish to use such
clusters effectively as probes of galaxy formation and evolution. This
contribution begins with a broad (and, by necessity, highly
incomplete) overview of what is known about the overall properties of
star cluster systems. It should then become clear that it is difficult
to answer the question as stated above, since there is no clear-cut
physical criterion that allows us to decide when a cluster should be
classified as `massive'. It is therefore useful to rephrase the
problem in a way that makes it more tractable. The remainder of the
current review will thus focus on three main themes: (i) the general
problem of cluster formation, (ii) dynamical evolution and (iii) the
shape of the initial cluster mass function (ICMF).

\section{Basic observational results}

\subsection{Open and globular clusters in the Milky Way}

In the context of this volume, it is appropriate to recall that the
first comprehensive discussion of the properties of star clusters was
given by Sir William Herschel in a series of papers published in the
\emph{Phil. Trans. R. Soc. London}. Herschel noted significant
differences in the visual appearances of clusters. He used the term
\emph{globular clusters} to describe the richest and most concentrated
of them (Herschel 1814). The term \emph{open} cluster emerged during
the early 20th century (Shapley 1916) as a common label for all
nonglobular clusters.

Originally, this classification was purely morphological, based simply
on the visual appearance of a cluster through a telescope or on a
photograph. Differences in spatial distribution, with the open
clusters concentrated near the Galactic plane and the GCs tending to
avoid it, were recognized early on (Shapley 1916; and references
therein). The open clusters are, in general, metal-rich with
metallicities similar to or even exceeding the solar value (Friel \ea
2002), while the Milky Way GC metallicity distribution is bimodal,
with both peaks at subsolar values (logarithmic iron abundance,
relative to solar, of [Fe/H]$\approx -1.5$ and $-0.5$ dex; Zinn
1985). In modern terms, these differences reflect the association of
the open clusters with the disc of our Galaxy and the GCs with the
spheroid (bulge/halo).

While the GCs are all ancient, with ages on the order of $10^{10}$
years and a spread of perhaps a few $\times 10^9$ years
(Mar\'{\i}n-Franch \ea 2009), the open clusters are mostly younger
than a few $\times 10^8$ years (Wielen 1971), although some older open
clusters are also known (Friel 1995). The lack of young GCs in the
halo and bulge can be attributed to a cessation of star formation in
these components long ago, but the field stars in the Galactic disc
have a continuous range of ages and open clusters are likely to have
formed there also in the distant past. The relative deficit of old
open clusters, therefore, illustrates that cluster \emph{dissolution}
is important.

The mass functions (MFs) of open and globular clusters are strikingly
different. The MF of young open clusters can be fitted by a power law,
$\rd N/\rd M \propto M^{-2}$, down to a few hundred M$_\odot$
(Elmegreen \& Efremov 1997; Piskunov \ea 2008). In contrast, the GC MF
is about flat at low masses, $\rd N/\rd M \sim$ constant for $M <
10^5$ M$_\odot$ (McLaughlin \& Pudritz 1996), while the high-mass end
can be fitted by a power law with slope $\sim -2$, as for the open
clusters. Unless the GCs were born with a different MF than young
clusters today, this flattening is another hint at the importance of
dynamical evolution.

Even the most massive open clusters identified in the Galactic disc
have masses below $\sim10^5$ M$_\odot$ (Davies \ea 2007; Brandner \ea
2008; Froebrich \ea 2009), an order of magnitude lower than the most
massive old GCs (Meylan \& Mayor 1986). Clearly, this difference
cannot easily be attributed to dynamical evolution, and seems all the
more puzzling given that the stellar mass of the Milky Way disc
exceeds that of the spheroid by about an order of magnitude (Dehnen \&
Binney 1998). We return to this issue below, but note here that GCs
and open clusters are subject to very different selection
biases. Extinction of optical light by interstellar dust in the
Galactic plane, combined with the high stellar density (`crowding')
along a given line of sight, strongly limits our ability to detect
distant open clusters. In fact, the discrepancy between `diameter
distances' (unaffected by extinction) and `photometric' distances of
open clusters led to one of the first quantitative estimates of the
amount of dust extinction in the Galactic plane (Trumpler
1930). Current catalogues of open clusters can only be considered
reasonably complete within $\sim1$ kpc of the sun (Lamers \ea 2005;
Piskunov \ea 2008). It is reasonable to assume that the most massive
and most luminous objects can be detected out to greater distances,
but the actual completeness of current surveys remains poorly
quantified. GCs are, instead, easier to find. Some may still remain
hidden in the plane, but most are found at higher Galactic latitudes,
where extinction and crowding are less severe, and the spatial
distribution of known GCs shows no tendency to concentrate near the
sun (e.g., Frenk \& White 1982).

\subsection{The Local Group}

The Magellanic Clouds are our nearest large extragalactic
neighbours. It has long been known that both Clouds, and the Large
Magellanic Cloud (LMC) in particular, are home to large numbers of
star clusters. Although the LMC is about a factor of ten less luminous
than the Milky Way (van den Bergh 1999), we can view it in its
entirety. The angular extent is about $5^\circ\times5^\circ$,
corresponding to about $5\times5$ kpc$^2$ (adopting a distance of
about 50 kpc), a significantly larger area than covered by
open-cluster surveys in the Milky Way.

The nomenclature used for LMC clusters has been a source of some
confusion. Many of the richest clusters were listed as globular when
discovered by John Herschel, while Shapley (1930) identified eight
`true' LMC GCs. However, all but one of these (NGC~1835) are now known
to be much younger than the Milky Way GCs. Many modern studies tend to
use old age as a defining criterion when identifying the GCs in the
Clouds (and elsewhere). By this metric, the LMC has about 13 GCs
(e.g., Schommer \ea 1992) while the SMC has one. This criterion works
well because of a large gap between ages of $\sim3\times10^9$ and
$\sim10^{10}$ years that naturally separates the old LMC GCs from
younger clusters (Rich \ea 2001).

The luminosity function (LF) of the old LMC clusters is similar to
that of Galactic GCs (Harris 1991), suggesting that the MFs are also
similar. The most massive young LMC clusters have masses of $\sim10^5$
M$_\odot$ (Fischer \ea 1992), somewhat exceeding the most massive
young clusters currently known in the Milky Way, but this might simply
be a size-of-sample effect because of the larger area surveyed and the
higher surface density of clusters in the LMC (Larsen 2002). There is
no evidence of significant differences in the actual MFs and LFs of
the young LMC and Milky Way clusters, with both following similar
power-law distributions (Elson \& Fall 1985; Hunter \ea 2003; de Grijs
\& Anders 2006).

An interesting contrast to the rich cluster systems of the Magellanic
Clouds is provided by the dwarf irregular galaxy IC~1613. In spite of
some ongoing star formation, this galaxy contains very few star
clusters in comparison with the Clouds, even when accounting for its
somewhat lower luminosity (van den Bergh 1979).

The two other Local Group spirals, M31 and M33, both host populations
of young and old star clusters that appear to be roughly equivalent to
the Milky Way open and globular clusters (Galleti \ea 2006; Sarajedini
\& Mancone 2007). The catalogues remain highly incomplete, however,
and the nature of many candidates remains to be verified. To date,
about 350 GCs have been confirmed in M31, while about 30 old, GC-like
objects have been identified in M33 (Schommer \ea 1991; Chandar \ea
2001; Huxor \ea 2009). Identification of clusters superimposed on the
discs is challenging because of crowding and confusion issues (Cohen
\ea 2005), so that information about the global properties of young
cluster populations in the discs of M31 and M33 is still relatively
scarce. However, both spirals host rich, young star cluster
populations similar to those observed in the LMC, with estimated
masses of up to $10^4 - 10^5$ M$_\odot$. The LFs and MFs are not well
constrained, but appear consistent with those in the Milky Way and the
LMC (Chandar \ea 2001; Sarajedini \& Mancone 2007; Caldwell \ea 2009).

\subsection{Beyond the Local Group}

Observations of extragalactic young clusters have been reviewed on
many previous occasions (e.g., Whitmore 2003; Larsen
2006). Identification of star clusters in star-forming galaxies beyond
the Local Group is challenging on the basis of ground-based
observations. In their study of young clusters in external galaxies,
Kennicutt \& Chu (1988) listed data for 14 galaxies, of which half are
members of the Local Group. The launch of the {\sl Hubble Space
Telescope (HST)} led to a revolution in the field, starting with the
discovery of a large number of bright, blue compact star clusters in
the galaxy NGC~1275 (Holtzman \ea 1992). With careful modelling of the
{\sl HST} point-spread function, a typical cluster with a half-light
radius of $\sim3$ pc remains recognizable as an extended object out to
distances of at least 40 Mpc (Harris 2009). This leads to a formidable
increase in the number of galaxies accessible to detailed study of
their cluster populations: Larsen (2006) lists 92 young systems for
which data were available as of 2004.

Many of the extragalactic systems that have been studied in detail are
starburst and merging galaxies, of which the best-studied case is
arguably the `Antennae' system, NGC~4038/4039. Like other ongoing
major gas-rich mergers, this pair is experiencing vigorous star
formation, including in a large number of luminous, compact star
clusters (Whitmore \ea 1999). Another well-studied system is the
nearby starburst M82, which also hosts many luminous young clusters,
although detailed analysis of their properties is hampered by heavy
extinction as the system is viewed nearly edge-on (O'Connell \ea 1995;
de Grijs \ea 2005; Smith \ea 2007). These cases are fairly typical of
the many systems that have been studied with the {\sl HST}. Where
cluster masses have been derived, they are often in the range $10^4 -
10^6$ M$_\odot$ or higher, comparable to the most massive old GCs
(Zhang \& Fall 1999; McCrady \& Graham 2007), with the lower end of
the range usually being set by detection limits.

An increasing amount of data for normal spiral galaxies have also
become available. Young clusters in the mass range $10^5 - 10^6$
M$_\odot$ have been found in some spirals (Larsen \& Richtler 2000,
2004), showing that such objects are not unique to starbursts and
interacting systems, although they may be more common there. It is
worth emphasizing that most spirals in which young cluster populations
have been studied are of Hubble type Sb or later, while little is
known about young clusters in Sa-type spirals.

The above discussion has concentrated on age and mass as the main
parameters characterizing star cluster properties. Another useful
parameter is the half-light radius, $R_\mathrm{h}$, which is expected
to remain approximately constant over the lifetime of a cluster
(Spitzer 1987). This is typically a few pc for both open and globular
clusters, independent of mass, although for GCs $R_\mathrm{h}$ tends
to correlate with Galactocentric distance and GCs as large as
$R_\mathrm{h}>20-30$ pc exist in the outer haloes of the Milky Way,
M31 and M33 (van den Bergh \ea 1991; Huxor \ea 2008, 2009). For very
massive clusters, there appears to be a more significant positive
correlation between cluster mass and size (Harris 2010). Unusually
extended ($R_\mathrm{h}>7$ pc) old clusters have also been found in
several S0-type galaxies (Larsen \& Brodie 2000; Hwang \& Lee 2006;
Peng \ea 2006). These `faint fuzzy' clusters (FFs) are distinctly
different from the outer-halo GCs in the Local Group spirals, since
they are clearly associated (kinematically and spatially) with the
\emph{discs} of their parent galaxies in at least a few cases (Burkert
\ea 2005). Their LFs also differ from that of normal old GCs by
showing no `turnover' near absolute visual magnitude $M_V\sim-7.5$
(current data do not constrain the LFs of FFs fainter than $M_V\sim-6$
mag). This may provide an important clue to the importance of
different disruption mechanisms in the discs of S0 galaxies, since
such extended clusters would be less affected by internal dynamical
evolution due to two-body relaxation, but more sensitive to external
shocks (Vesperini 2010). It remains largely unknown whether FFs form
through a special channel that operates predominantly in the discs of
S0 galaxies, or if this environment is particularly favourable for
their survival. An interesting possibility is that FFs may have formed
by the merger of smaller subunits (Fellhauer \& Kroupa 2005; Burkert
\ea 2005), perhaps similarly to the star-cluster complexes observed in
some galaxies with ongoing cluster formation (Bastian \ea 2005$a$).

\section{Formation of clusters}
\label{sec:formation}

The problem of cluster formation is intimately linked to that of star
formation. There are many excellent reviews on this topic (e.g., Lada
\& Lada 2003; Mac Low \& Klessen 2004; McKee \& Ostriker 2007) and the
discussion in this section will concentrate on a few issues of
relevance to the global properties of cluster populations. More
detailed discussion can be found elsewhere in this volume (Clarke
2010; Lada 2010).

\subsection{From giant molecular clouds to (embedded) star clusters}

Star formation is closely associated with dense molecular gas and
clusters are observed to form deeply embedded within giant molecular
clouds (GMCs). The structure of these GMCs is self-similar on a wide
range of scales down to individual protostellar cores, which can be
grouped into cluster-forming clumps (Williams \ea 2000; McKee \&
Ostriker 2007). The GMCs are themselves part of a larger hierarchy of
structure in the interstellar medium (Elmegreen \& Falgarone 1996;
Elmegreen 2007), and tend to be organized into giant molecular
\emph{complexes} (Wilson \ea 2003), which are located along the spiral
arms in spiral galaxies (Vogel \ea 1988). Averaged over an entire GMC,
the density of molecular gas is low ($n_\mathrm{H} \sim10^2-10^3$
cm$^{-3}$) and stars only form in the densest regions ($n_\mathrm{H} >
10^5$ cm$^{-3}$). Globally, star formation is therefore an inefficient
process, and only a few percent of the mass of a given GMC is
converted into stars before the cloud is dispersed (Williams \& McKee
1997). The GMC-wide star-formation efficiency,
$\epsilon_\mathrm{GMC}$, should not be confused with the \emph{local}
star-formation efficiency, $\epsilon_\mathrm{cl}$, within the
cluster-forming clumps, which must be at least 20--30\% to produce a
bound cluster (section~\ref{sec:dynev}).

An interesting question is how the MF of GMCs is related to that of
the clusters forming within them. The GMC mass function in the Milky
Way has a characteristic upper mass of $\approx 6\times10^6$
M$_\odot$. Below this mass, it can be approximated by a power law,
$\rd N/\rd M \propto M^{-1.7}$, but it declines steeply at higher
masses (Williams \& McKee 1997). For $\epsilon_\mathrm{GMC} \sim 5$\%,
the upper GMC mass would correspond to a cluster mass of
$\sim3\times10^5$ M$_\odot$, although the ICMF is unlikely to be a
simple scaled-down version of the GMC MF, since a single GMC may form
more than one cluster (e.g., Kumar \ea 2004). The mass spectrum of
clumps within GMCs may be more relevant. This appears to follow a
similar power law as that of the GMCs, $\rd N/\rd M \propto
M^{\alpha}$, with $\alpha\approx-2$ but with a large uncertainty (Mac
Low \& Klessen 2004). Nevertheless, an upper limit of a few $\times
10^5$ M$_\odot$ is consistent with other constraints on the ICMF in
spiral galaxies (section~\ref{sec:lfmf}; Larsen 2009).

It is difficult to see how the most massive clusters observed in some
external galaxies, with masses of $10^6$ M$_\odot$ or higher, could
form from Milky Way-like GMCs. For a $10^6$ M$_\odot$ cluster with a
half-mass radius of 3 pc, the \emph{current} mean density within the
half-mass radius corresponds to $\langle n_\mathrm{H} \rangle =
2\times10^5$ cm$^{-3}$. This is a strict lower limit to the density of
the gas from which the cluster must have formed, since
$\epsilon_\mathrm{cl} < 1$ and clusters expand following gas expulsion
(Goodwin \& Bastian 2006; Scheepmaker \ea 2007; Bastian \ea 2008;
Pfalzner 2009). The formation of the most massive clusters requires
collecting the amount of gas typical of a massive Galactic GMC within
a volume only a few pc across, essentially turning such a cloud into
one big clump. A likely key element to understanding how such dense,
massive clumps can exist is the high gas densities in starburst
galaxies. This may allow denser and more massive GMCs to condense
(Escala \& Larson 2008), perhaps further aided by shock compression in
mergers (Jog \& Solomon 1992; Ashman \& Zepf 2001). There is
observational evidence that GMCs in M82 are indeed compressed to
higher densities than their Milky Way counterparts by the ambient
pressure (Keto \ea 2005), so that in these clouds massive clusters may
form at high $\epsilon_\mathrm{GMC}$. In even more extreme
environments, such as in ultraluminous infrared galaxies, GMCs may be
both denser and more massive (Murray \ea 2009), consistent with the
presence of clusters with $M > 10^7$ M$_\odot$ in some merger remnants
(Maraston \ea 2004; Bastian \ea 2006).

\subsection{The embedded phase}

Observations of the embedded phase are challenging (cf. Lada 2010)
because of its short duration, combined with the high gas column
densities in GMCs ($N_\mathrm{H} \sim 1.5\times10^{22}$ cm$^{-2}$;
McKee \& Ostriker 2007) and corresponding large amounts of dust
extinction ($A_V\sim8$ mag). The duration of this phase is uncertain,
since it is difficult to determine the age of (unresolved) embedded
clusters and the assumption of a single age may be questionable in the
first place. Upper limits may be set by age dating clusters that have
already become optically visible. The R136 cluster in the LMC contains
some stars as young as 1--2 Myr which have mean extinctions of only
$A_V\sim1.2$ mag (Massey \& Hunter 1998). Whitmore \& Zhang (2002)
find optically visible counterparts for about three quarters of the
brightest radio-continuum sources in the Antennae. Among these,
clusters older than $\sim 2.5$ Myr all have low extinctions of $0.5 <
A_V < 2.5$ mag. Similarly, Reines \ea (2008) find that radio-detected
clusters in NGC~4449 with ages in the range of 3--5 Myr already have
low extinctions, $A_V=0.5-1.5$ mag. The bright cluster NGC~1569--A,
with an age of $\approx5$ Myr (Origlia \ea 2001; Maoz \ea 2001) is
essentially free of dust extinction. From these examples, it is clear
that the embedded phase lasts at most a few $\times 10^6$ years.

What physical mechanism is responsible for expelling the gas? The
ionizing radiation from massive stars will produce an H{\sc ii}
region, but the thermal pressure in the ionized gas may be
insufficient to overcome self-gravity for clusters more massive than
$\sim10^5$ M$_\odot$ (Kroupa \& Boily 2002).  Various alternatives are
discussed by Krumholz \& Matzner (2009). Supernovae could easily
provide enough energy but appear after several $10^6$ years, probably
too late to explain the observed short duration of the embedded phase.
Another candidate is winds from massive stars, which have velocities
exceeding 1000 km s$^{-1}$ and also provide more than sufficient
energy to unbind even massive ($M>10^6$ M$_\odot$) clusters. However,
the efficiency of such winds may be low. Krumholz \& Matzner finally
conclude that \emph{radiation pressure} from massive stars is most
likely the dominant gas-evacuation mechanism in massive clusters.

\section{Dynamical evolution}
\label{sec:dynev}

\subsection{Early dynamical evolution: `infant mortality'}

The formation of an embedded cluster does not guarantee that it will
remain bound after gas expulsion (see also the discussion in Lada
2010). If the stars and gas are initially in virial equilibrium, the
velocity dispersion of the stars will be too high to match the
shallower potential once the gas is expelled. If gas expulsion happens
instantaneously and $\epsilon_\mathrm{cl}<50$\%, the cluster will
dissolve completely, independent of the initial mass (Hills 1980). In
practice, cluster expansion does not occur instantly so the stars have
some time to adjust to the new potential, while gas expulsion is
probably not instantaneous either. Simulations suggest that at least
some fraction of the stars may remain bound for lower star-formation
efficiencies, perhaps as low as $\epsilon_\mathrm{cl} \approx20-30$\%
(Boily \& Kroupa 2003; Goodwin 1997; Baumgardt \& Kroupa 2007). The
timescale for the cluster to settle into a new equilibrium may be as
long as several $\times 10^7$ years (Goodwin \& Bastian 2006).

This picture is supported by the observation that about 95\% of
clusters formed in Milky Way GMCs dissolve in less than $10^8$ years
(Lada \& Lada 2003). However, the universality of this
`infant-mortality' (IM) fraction is poorly quantified. The term
`mass-independent disruption' (MID) is also sometimes used to
distinguish this process from the secular, mass-dependent dissolution
that occurs on longer timescales and which will be discussed
below. The age distribution of mass-limited cluster samples in the
Antennae galaxies is approximately $\rd N/\rd\tau \approx \tau^{-1}$
for ages ($\tau$) of up to $10^8-10^9$ yr (Fall \ea 2005), suggesting
that $\sim80-90$\% of the clusters disappear per decade in age,
independent of mass (Whitmore \ea 2007). However, over such a large
age range, it is unlikely that disruption can still be attributed to
gas expulsion. One difficulty with estimating the disruption
parameters in an interacting system like the Antennae is that the
star- and cluster-formation rates may not have been constant in the
past. Bastian \ea (2009) find that the age distribution of clusters in
the Antennae can also be fit by a model in which the cluster-formation
rate has increased over the past few $\times 10^8$ years (as suggested
by simulations of the ongoing interaction), with MID required only
over a period of $\sim10^7$ years. Evidence of a large IM fraction
($\sim70$\%) has also been claimed in M51 (Bastian \ea 2005$b$), but
this may be partly due to age-dating artifacts around $10^7$ years
(Gieles 2009). In the SMC, Chandar \ea (2006) derive an age
distribution of $\rd N/\rd \tau \sim \tau^{-0.85}$ for $\tau<10^9$
years, similar to that seen in the Antennae, but Gieles \ea (2007)
argue that this may be caused by fading below the detection limit at
old ages and instead conclude that no significant MID is needed to
explain the age distribution of SMC clusters (see also de Grijs \&
Goodwin 2008 for evidence supporting the latter scenario).

\subsection{The cluster-formation efficiency}

In starburst and merger systems, young clusters often account for a
large fraction (10--20\%) of the total blue/ultraviolet flux (Tremonti
\ea 2001; Meurer \ea 1995; Zepf \ea 1999; Fall \ea 2005), consistent
with essentially all stars forming in clusters (see also Johnson \ea
2009). In the Milky Way, the majority of stars form in embedded
clusters (Lada \& Lada 2003). However, most stars eventually end up
belonging to the field. Observationally, it is difficult to tell
whether \emph{all} stars were born in clusters, with a large fraction
dispersing almost immediately, or whether some stars were born in
genuinely dispersed mode.

For practical purposes, one may still define a cluster-formation
`efficiency' as the ratio of the number of stars that end up in
clusters relative to the field. Since this ratio will depend on age,
ideally some age range should be specified. For optically visible
clusters, Bastian (2008) finds a constant efficiency of
$\Gamma\sim8$\% in galaxies spanning six orders of magnitude in
star-formation rate (SFR). On the other hand, the fraction of the
total $U$-band light in galaxies originating from clusters correlates
with the area-normalized SFR of the parent galaxy, ranging from well
below 1\% in quiescent systems to the high numbers found in starbursts
(Larsen \& Richtler 2000). The near-absence of clusters in IC~1613,
and the nonuniversal GC \emph{specific frequency} (number of GCs per
unit host-galaxy luminosity; Harris 1991, 2010), are other hints that
variations in $\Gamma$ may exist.

Even if a single indicator of star formation (e.g., the far-infrared
or ultraviolet luminosity) is used, different systematic errors will
affect galaxies differing in dust content, metallicity, ratio of
current to past SFR and other parameters (Kennicutt 1998). Such
systematic errors on the parent-galaxy SFRs are likely different than
those associated with the cluster-formation rates, which are typically
inferred from direct observations of cluster populations but still
subject to uncertainties due to disruption, potential confusion with
other objects and completeness effects. Consequently, $\Gamma$ remains
difficult to constrain.

\subsection{Secular evolution}
\label{sec:clev}

Clusters that survive IM will continue to evolve dynamically on longer
timescales as a result of internal two-body relaxation, external
shocks and mass loss due to stellar evolution (Vesperini 2010). This
`secular evolution' will lead to the gradual evaporation of any star
cluster and, eventually, its total dissolution. Two-body relaxation
causes the velocities of the stars in the cluster to approach a
Maxwellian distribution and stars with velocities above the escape
velocity will gradually evaporate from the cluster. The two-body
relaxation time scales as $t_\mathrm{rel} \propto \sqrt{M} \,
R_\mathrm{h}^{3/2}$, but the actual evaporation time, $t_\mathrm{ev}$,
may scale nonlinearly with $t_\mathrm{rel}$ (Baumgardt \& Makino
2003). In addition, external shocks will lead to dissolution on a
timescale $t_\mathrm{sh} \propto M R_\mathrm{h}^{-3}$ (Spitzer
1987). Shocks may be due to encounters with spiral arms, GMCs or, for
GCs on eccentric orbits, passages through the galactic disc or near
the bulge. Finally, stellar evolution causes mass loss as stars are
turned into much less massive remnants. Over a 10 Gyr time span, about
one third of the initial cluster mass is lost this way (e.g., Bruzual
\& Charlot 2003).

For many practical purposes, secular dissolution may be conveniently
characterized by a single \emph{dissolution timescale}
$t_\mathrm{dis}$, such that mass is lost at a rate $\left(\rd M/\rd
t\right)_\mathrm{dis} = -M/t_\mathrm{dis}$. The dissolution timescale
may be parameterized as $t_\mathrm{dis} = t_4 (M/10^4
\mathrm{M}_\odot)^\gamma$, where $t_4\sim10^9$ years and
$\gamma\sim0.65$ for clusters in the solar neighbourhood (Lamers \ea
2005). The relatively larger number of old clusters in the Magellanic
Clouds suggests that disruption is less efficient there (Elson \& Fall
1985; Girardi \ea 1995; Hodge 1987; de Grijs \& Anders 2006), while
the disruption timescale in the central regions of M51 appears much
shorter than in the solar neighbourhood (Boutloukos \& Lamers
2003). This may be caused by different densities of GMCs in these
environments (Wielen 1985; Terlevich 1987; Gieles \ea 2006$b$).

Since low-mass clusters disrupt faster, the MF will flatten over time
and, given sufficient time, ($\tau \gg t_\mathrm{dis}$), the MF will
tend towards $\rd N/\rd M \propto M^{\gamma-1}$. Hence, observations
of the MF in cluster systems of different ages can potentially be used
to constrain the disruption law. Good fits to the GC LF in the Milky
Way and other galaxies can indeed be obtained if cluster MFs similar
to those observed in young cluster systems in merging galaxies are
evolved with $t_4\sim10^8-10^9$ years and $\gamma=0.7-1$ (Fall \&
Zhang 2001; Jord{\'a}n \ea 2007; McLaughlin \& Fall 2008; Kruijssen \&
Portegies Zwart 2009). However, radial variations in the GC LF are
expected, since both shocks and tidal fields will be weaker at large
galactocentric radii. The fact that such variations are not observed
in old GC systems is a potential difficulty (Vesperini \ea
2003). Interestingly, observations of the intermediate-age
($\sim3\times10^9$ years) merger remnant NGC~1316 \emph{do} show a
radial variation in the MF with a higher turnover mass near the centre
(Goudfrooij \ea 2004).

\section{The initial cluster mass function}
\label{sec:lfmf}

The basic stellar dynamical mechanisms responsible for the evolution
of the MF are the same for all clusters, even though the relative
importance of different mechanisms (e.g., two-body relaxation versus
shocks) may differ. However, the ICMF on which these mechanisms
operate might still vary with environment. Once observational
selection effects are accounted for, it is relatively straightforward
to derive the \emph{luminosity} function of a cluster sample, assuming
the distance is known. However, the interpretation of LFs in terms of
the physically more fundamental MF is complicated by the fact that not
all clusters have the same mass-to-light ratio, $\Upsilon$. A direct
conversion from LF to MF is therefore not usually
possible. Dissolution makes reconstruction of the ICMF even more
challenging.

For samples that are large enough to derive statistically meaningful
MFs, the only practical approach is to derive the \emph{ages} of
individual clusters and then make use of `simple stellar population'
(SSP) models that tabulate $\Upsilon$ versus age to convert the
luminosities to masses (e.g., Bruzual 2010). Ages are typically
derived from integrated broad-band colours (e.g., $UBVRI$). However,
these are sensitive to both age and other parameters such as
extinction and metallicity. In principle, the use of multiple filters
allows to solve for all of these parameters, although using different
SSP models and fitting methods can still lead to large systematic
differences in the derived ages (at least a factor of 2; de Grijs \ea
2005; Scheepmaker \ea 2009). In addition, stochastic effects due to
the finite number of stars in a cluster can lead to large departures
from the predicted colours, especially for low-mass, young clusters
(Girardi \ea 1995; Bruzual \& Charlot 2003; Cervi{\~n}o \& Luridiana
2006; Ma{\'{\i}}z Apell{\'a}niz 2009). The requirement for multiple
filters is also costly in terms of observing time, especially at blue
and ultraviolet wavelengths where detectors tend to be less sensitive.

\begin{figure}
\centerline{\includegraphics[width=75mm]{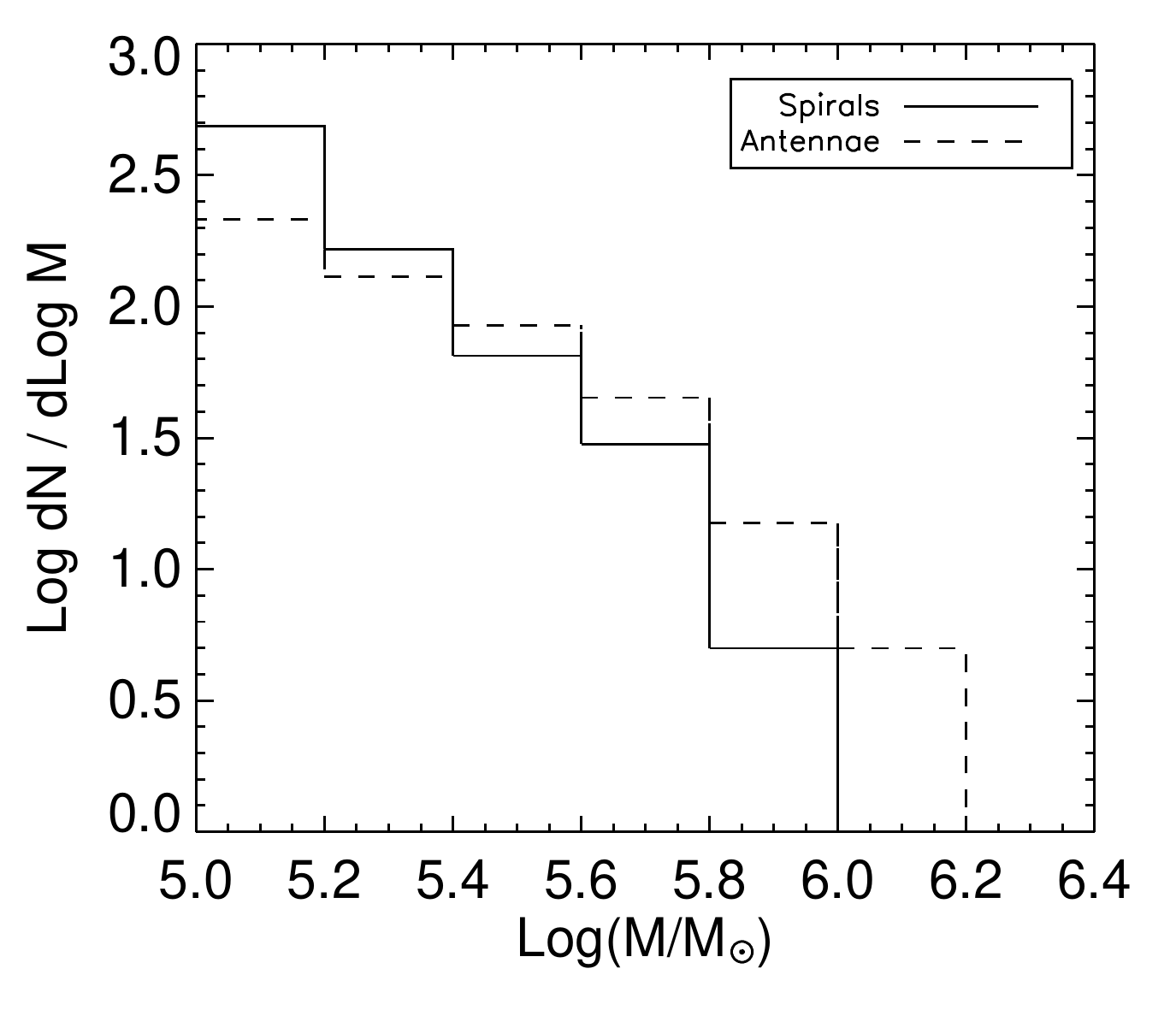} }
\caption{\label{fig:mfcmp}Mass functions for young ($<2\times10^8$
years) clusters in spiral galaxies (Larsen 2009) and the Antennae
system (Whitmore \ea 1999).}
\end{figure}

Determinations of the MF are only available for a few young cluster
systems. Generally, they are well-represented by power laws, $\rd N /
\rd M \propto M^{\alpha}$, with slopes of $\alpha \sim -2.0$, but the
mass ranges over which these slopes are derived vary considerably (see
Larsen 2009 for references). Figure~\ref{fig:mfcmp} shows a comparison
of the MFs for young clusters in spiral galaxies (taken from Larsen
2009) and the Antennae system (Zhang \& Fall 1999). Ground-based data
for 17 spirals have been combined to improve statistics, although 75\%
of the clusters belong to the two most cluster-rich galaxies, NGC~5236
and NGC~6946. The combination of data for many spirals is justified
since the MFs in different subsamples are statistically
indistinguishable (Larsen 2009). The Antennae clusters span the range
25$\times10^6 <\tau< 160\times10^6$ years, while clusters younger than
$2\times10^8$ years are included for the spirals. Completeness limits
restrict the useful mass range to $10^5 < M/\mathrm{M}_\odot$, but the
figure already hints at differences between the MFs, with relatively
more high-mass clusters in the Antennae. This is confirmed by a
Kolmogorov--Smirnov test, which yields a very small probability
($P=0.00032$) that the two samples are drawn from the same parent
distribution.

The high-mass end of the MF in old GC systems is well approximated by
a Schechter (1976) function,
\begin{equation}
  \frac{\rd N}{\rd M} \propto \left( \frac{M}{M_\mathrm{c}}
  \right)^{\alpha} \exp \left( - \frac{M}{M_\mathrm{c}}\right),
  \label{eq:schechter}
\end{equation}
with cutoff mass $M_\mathrm{c} >$ several $\times 10^6$ M$_\odot$
(Burkert \& Smith 2000; McLaughlin \& Pudritz 1996; Jord{\'a}n \ea
2007). This should not be confused with the turnover at $\sim10^5$
M$_\odot$ which is most likely a result of dynamical evolution. A
Schechter-function fit to Antennae data in figure~\ref{fig:mfcmp}
yields $M_\mathrm{c} = (1.7\pm0.7)\times10^6$ M$_\odot$ for fixed
$\alpha=-2$ (see also Jord{\'a}n \ea 2007), although a uniform power
law with no truncation is also consistent with the data (Whitmore \ea
2007). A fit to the spiral data instead gives $M_\mathrm{c} =
(2.1\pm0.4)\times10^5$ M$_\odot$, and a uniform $\alpha=-2$ power law
is ruled out at high confidence level (Larsen 2009). This again
indicates a dearth of high-mass clusters in spirals, compared to the
Antennae.

Although the MF and LF are not the same, they are of course
related. If $\psi_\mathrm{i}(M_\mathrm{i})$ is the ICMF, normalized to
unit mass over some range of \emph{initial} cluster mass
$M_\mathrm{low} < M_\mathrm{i} < M_\mathrm{up}$, the LF is (Larsen
2009)
\begin{equation}
 \frac{\rd N}{\rd L} = \int_{\tau_\mathrm{min}}^{\tau_\mathrm{max}}
  \psi_\mathrm{i}\left[M_\mathrm{i}(L,\tau)\right] \times \frac{\rd
  M_\mathrm{ i}}{\rd M} \times \Upsilon \times \Gamma \times
  \mathrm{SFR} \times f_\mathrm{surv}(\tau) \, \rd \tau .
  \label{eq:lf}
\end{equation}

The rate of star formation in clusters is expressed as $\Gamma \times
\mathrm{SFR}$. Infant mortality is formally included as a
mass-independent survival fraction, $f_\mathrm{surv} =
(\tau/\tau_0)^{\log(1-\mathrm{IMR})}$ where $\tau_0$ marks the onset
of IM and IMR is the fraction of clusters lost per decade in age. For
$\tau<\tau_0$, $f_\mathrm{surv} = 1$ and IM may be switched off at
some time $\tau_\mathrm{IM,max}$ after which $f_\mathrm{surv}$ is
constant. The initial cluster mass $M_\mathrm{i}$ is related to the
current mass $M = \Upsilon L$ through the assumed disruption law. If
there is no disruption and the ICMF is a uniform power law, $\psi(M)
\propto M^\alpha$, then it follows from equation~(\ref{eq:lf}) that
the LF is a power law with the same slope. In general, however, the
shape of the LF will differ from that of the underlying MF because of
disruption and the age-dependent mass-to-light ratio.

\begin{figure}
\includegraphics[width=125mm]{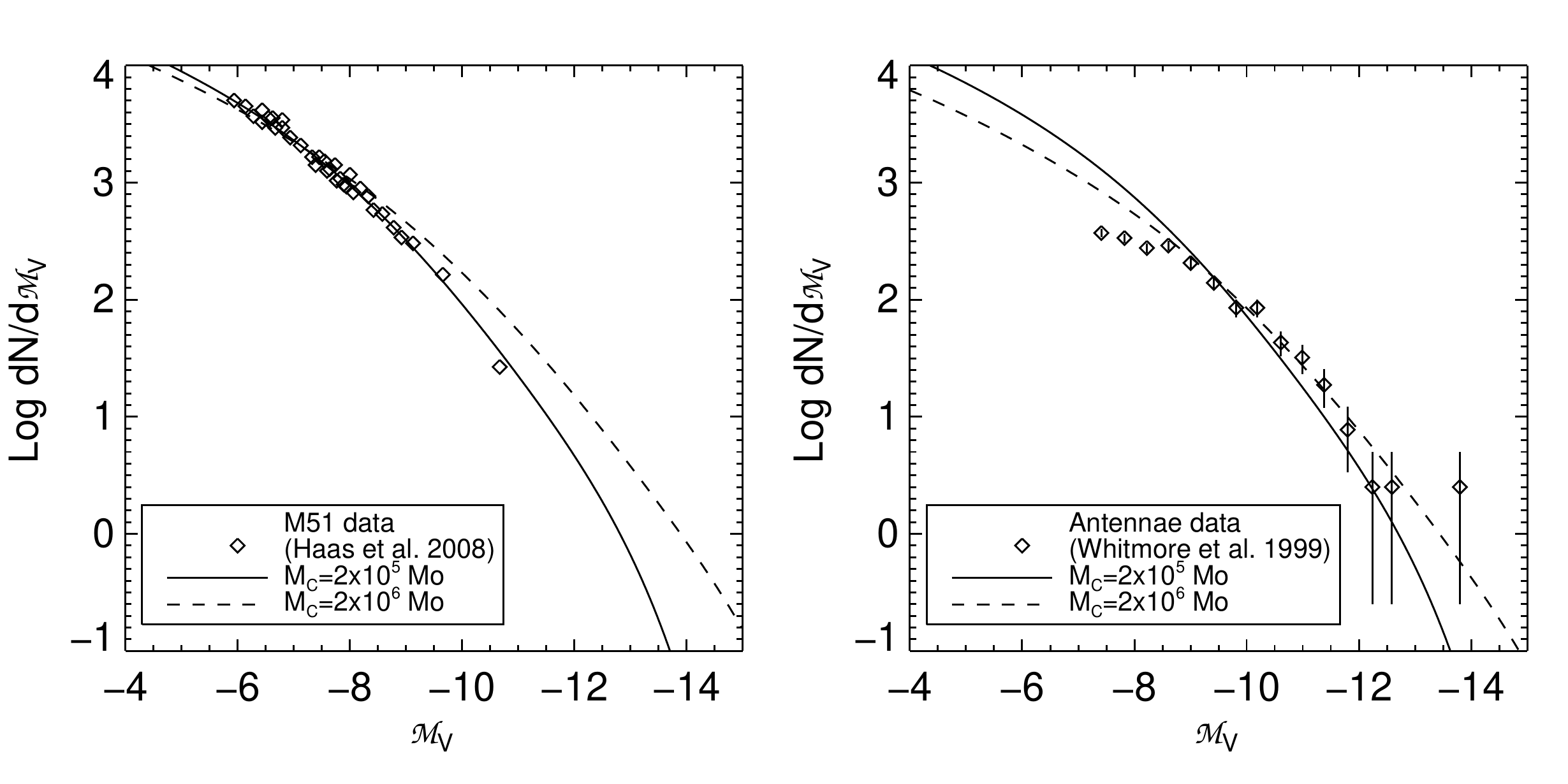} 
\caption{\label{fig:lfs}Luminosity functions of clusters in M51 (left)
and the Antennae galaxies (right). Also shown are model LFs assuming
Schechter ICMFs with $M_\mathrm{c}=2\times10^5$ and $2\times10^6$
M$_\odot$ (solid and dashed lines, respectively), scaled to match the
data. See text for details.}
\end{figure}

In figure~\ref{fig:lfs}, the LFs of clusters in the spiral galaxy M51
(Haas \ea 2008) and the Antennae system (Whitmore \ea 1999) are
compared with model LFs for Schechter ICMFs with $M_\mathrm{c} =
2\times10^5$ (solid line) and $2\times10^6$ M$_\odot$ (dashed
line). In both cases, an IMR of 80\% for ages (5--30)$\times10^6$
years is assumed, and secular dissolution is modelled using
$t_4=5\times10^8$ years and $\gamma=0.65$. SFRs are assumed constant,
but the model LFs have been shifted vertically to match the data. The
relatively small number of clusters in the Antennae compared to M51
originates from using the Whitmore \ea (1999) `PC: cluster-rich
region' sample, which covers only a small fraction of the full galaxy
pair (panel d in their figure 11). It is clear that the
$M_\mathrm{c}=2\times10^5$ M$_\odot$ model LF matches the M51 data
quite well, while the $M_\mathrm{c}=2\times10^6$ M$_\odot$ LF is too
shallow at the bright end. For the Antennae, however, the
$M_\mathrm{c}=2\times10^6$ M$_\odot$ LF provides a better fit. The
Antennae LF flattens further below $M_V\sim-8$ mag, but Whitmore \ea
(1999) indicate that the selection of cluster candidates becomes less
reliable below this limit. M51 is perhaps not the most typical spiral,
since it is mildly interacting with a nearby companion. However, the
LF comparison suggests that the MF there is more similar to that in
noninteracting spirals than in the merging Antennae galaxies.

Figure~\ref{fig:lfs} shows that the LF is expected (and observed) to
steepen towards the bright end, so that a single power law generally
provides a poor fit over an extended magnitude range. This steepening
has been noted in other data sets, including the merger NGC~3256 (Zepf
\ea 1999) and various spiral galaxies (Larsen 2002; Dolphin \&
Kennicutt 2002). Typical power-law slopes are between $-2.0$ and
$-2.5$. From similar modelling of the LF, Gieles \ea (2006$a$)
inferred a truncation of the MF around $M_\mathrm{up} \sim 10^5$
M$_\odot$ in M51 and another spiral, NGC~6946, and it was suggested
already by Zhang \& Fall (1999) that the steepening of the Antennae LF
brighter than $M_V\sim-10$ mag might indicate truncation of the MF
around $10^6$ M$_\odot$. Clearly, the LF has some diagnostic power,
although not without relying on model assumptions.

\subsection{A few notes on size-of-sample effects}

\begin{figure}
\includegraphics[width=65mm]{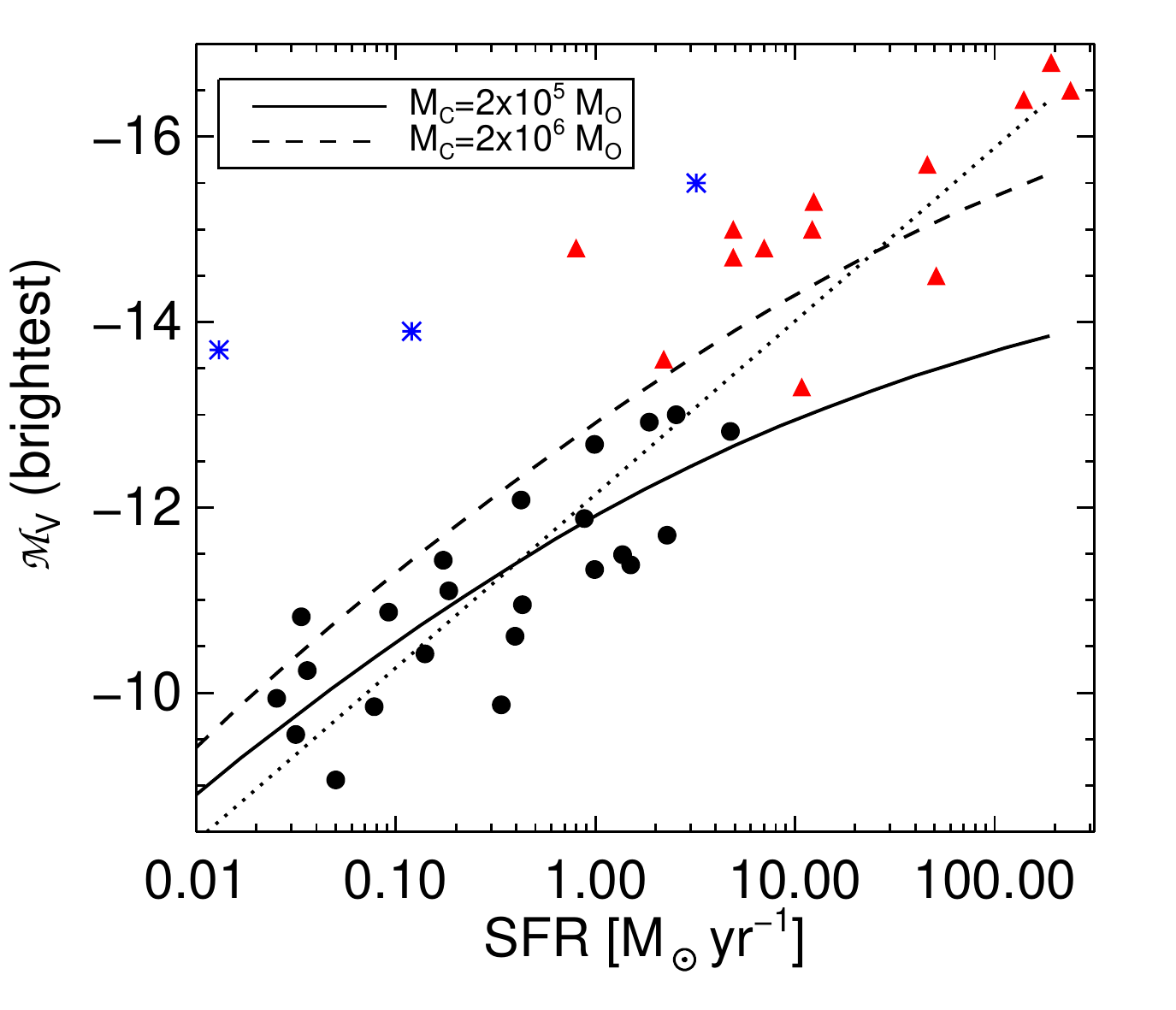}
\includegraphics[width=65mm]{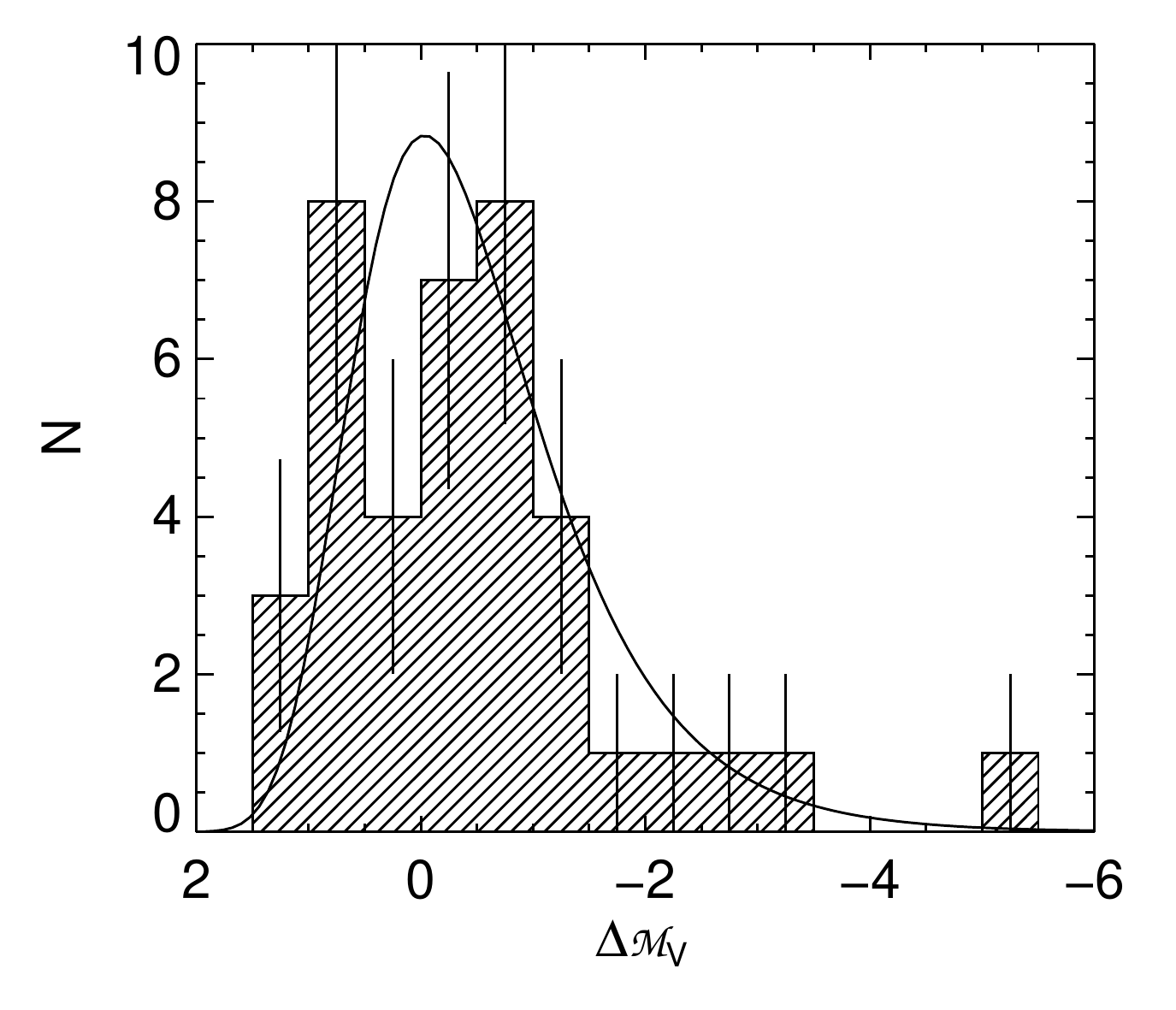}
\caption{\label{fig:lmax_sfr}{\it (left)} $M_V$ of the brightest
clusters in galaxies versus the galaxy-wide SFR. Circles: spiral
galaxies, triangles: mergers/interacting galaxies, asterisks:
others. Random sampling relations for $M_\mathrm{c} = 2\times10^5$ and
$2\times10^6$ M$_\odot$ are also shown (see text for details). The
dotted line is the best fit from Weidner \ea (2004). {\it (right)}
Distribution of differences, $\Delta M_V$, between the observed
brightest magnitude $M_V$(brightest) and the Weidner \ea fit. The
solid curve is the $\Psi$ distribution based on
equation~(\ref{eq:psil}).}
\end{figure}

It is well known that a tight correlation exists between the
luminosity of the \emph{brightest} cluster, $L_\mathrm{max}$, in a
galaxy and the total number of clusters or the overall SFR (Billett
\ea 2002; Larsen 2002; Whitmore 2003). An updated version of the
$L_\mathrm{max}$--SFR relation is shown in the left-hand panel of
figure~\ref{fig:lmax_sfr}, with spiral galaxies shown as filled
circles and interacting systems as triangles (Larsen 2002; Bastian
2008). A few dwarf galaxies are indicated by asterisks. The latter are
well-known outliers in this context that have been discussed
extensively in the literature (Billett \ea 2002; Larsen 2002; Whitmore
\ea 2007; Bastian 2008).

This relation is just the type of effect that is expected if cluster
luminosities are drawn at random from a LF which decreases towards the
bright end. It does not lead trivially to the conclusion that there is
a physical correlation of $M_\mathrm{c}$ versus SFR. Statistically,
the luminosity of the brightest cluster may be estimated by solving
$\ln 2 \sim 0.7 = \int_{L_\mathrm{max,med}}^\infty (\rd N/\rd L) \,
\rd L$, where the constant on the left-hand side is chosen such that
$L_\mathrm{max,med}$ is the median luminosity of the brightest
cluster. For a power-law LF, this leads to the relation
$L_\mathrm{max,med} \propto N^{-1/(\alpha+1)}$, where $N$ is the
number of clusters brighter than some minimum luminosity,
$L_\mathrm{min}$. The dotted line in figure~\ref{fig:lmax_sfr} is the
fit $L_\mathrm{max} \propto \mathrm{SFR}^{0.748}$ (Weidner \ea 2004),
which implies $\alpha\sim-2.3$ if $N \propto$ SFR. This is again
consistent with the bright-end slope of the LF being steeper than the
low-mass end of the ICMF (see also Whitmore \ea 2007).

If $L_\mathrm{max}$ is determined by random sampling, it is not a
single number for a given $N$ but a random variable. For a power-law
LF, the mean number of clusters brighter than some luminosity
$L_\mathrm{max}$ will be $\mu_\mathrm{b} = N
(L_\mathrm{max}/L_\mathrm{min})^{(1+\alpha)}$.  The probability that
there are no clusters brighter than a particular $L_\mathrm{max}$ is
then $P_\mathrm{b} = \re^{-\mu_\mathrm{b}}$. It follows that the
\emph{distribution} of brightest-cluster luminosities,
$\Psi(L_\mathrm{max}) = \rd P_\mathrm{b}/\rd L_\mathrm{max}$, is
\begin{equation}
 \Psi(L_\mathrm{max}) = -\frac{(1+\alpha)}{L_\mathrm{max}}
 \left(L_\mathrm{max}/L_\mathrm{ref}\right)^{1+\alpha}
 \exp\left(-\left[L_\mathrm{max}/L_\mathrm{ref}\right]^{1+\alpha}\right)
  \label{eq:psil}
\end{equation}
for $L_\mathrm{ref} = L_\mathrm{min} N^{-1/(\alpha+1)}$ (for a
slightly different approach, see Maschberger \& Clarke 2008). The tail
of this distribution approaches a power law with the same slope
$\alpha$ as the LF itself at high luminosities. It can be shown that
$L_\mathrm{ref}$ is the mode of the $\log L_\mathrm{max}$ distribution
and is related to the median of $\Psi(L_\mathrm{max})$ as
$L_\mathrm{max,med}/L_\mathrm{ref} = (\ln 2)^{1/(1+\alpha)}$. The
right-hand panel shows the distribution of residuals from the Weidner
\ea (2004) fit compared to equation~(\ref{eq:psil}). As found in
previous studies, the scatter around the fit is largely consistent
with random sampling (Larsen 2002; Whitmore \ea 2007), although it may
be significant that the largest residuals are mostly due to dwarfs. In
particular, further steepening of the LF towards the bright end would
make such outliers less likely.

For Schechter-like ICMFs, the LF is not expected to be a single power
law. Is this still consistent with the observed $L_\mathrm{max}$--SFR
relation? The solid and dashed lines in figure~\ref{fig:lmax_sfr} show
the $L_\mathrm{max,med}$ relations for the model LFs in
figure~\ref{fig:lfs}. We have further assumed $\Gamma = 0.50$, so that
about 14\% of all stars formed are still in clusters after
$3\times10^7$ years, at the end of the IM phase. Again, a single
$M_\mathrm{c}=2\times10^5$ M$_\odot$ Schechter ICMF fits all spiral
data quite well, but fails to reproduce the interacting systems. These
clearly require a higher upper-mass limit, and are instead well fit by
the $M_\mathrm{c}=2\times10^6$ M$_\odot$ model. The detailed model
parameters are poorly constrained and other combinations of $\Gamma$
and IM provide equally good fits. However, the observed
$L_\mathrm{max}$--SFR relation is consistent with other constraints on
the ICMF discussed in previous sections. It neither requires
$M_\mathrm{c}$ to scale in a simple way with the SFR, nor that the LF
and ICMF are completely untruncated, uniform power laws. The fact that
the $L_\mathrm{max}$--SFR relation holds over such a relatively large
dynamic range, even for Schechter-like ICMFs, is in part due to the
fact that the age of the brightest cluster (and hence $\Upsilon$) will
be a decreasing function of $N$ (or SFR) for fixed $M_\mathrm{c}$, so
that $L_\mathrm{max}$ is sensitive to size-of-sample effects even if
the MF is sampled up to near $M_\mathrm{c}$ (Larsen 2009).

\section{Concluding remarks}

The use of star clusters as tracers of extragalactic stellar
populations relies on a close link between star formation in general
and cluster formation. While several recent studies have converged on
a fraction of $\sim10$\% of stars forming in clusters that remain
bound for at least a few $\times10^7$ years, the ratio of clusters to
field stars varies enormously in old GC systems, both from galaxy to
galaxy and within galaxies (Harris 1991; Harris 2010). It is hard to
rule out that cluster dissolution is partly responsible for these
differences, particularly if it is mass independent (Whitmore \ea
2007), but so far there is no robust way to predict what fraction of
GCs may have been lost over a Hubble time in this way. While the
\emph{timing} of, e.g., a major burst of star formation may be
inferred from a peak in the cluster age distribution, the
\emph{strength} of such a burst remains much more poorly constrained.

Apart from differences in the formation efficiency of clusters, the
mass spectrum may also vary with environment. There are hints that the
ICMF in quiescent discs may be less top heavy than in violent
starbursts, so that \emph{massive} clusters predominantly trace the
latter. Although GC formation is no longer viewed as `special', this
suggests that GCs formed under conditions that are more similar to
those in present-day starbursts than in discs. Low-mass clusters,
formed under quiescent conditions long ago, may have dissolved by now.

In the coming years, observations of molecular gas in external
galaxies with the {\sl Atacama Large Millimeter Array} are likely to
provide a tremendous boost in our understanding of cluster formation
under conditions that differ from those in local star-forming
regions. The newly refurbished {\sl HST} is more capable than ever,
and will allow more detailed constraints on the age and mass
distributions of clusters in different galaxies, so that the role of
environment in determining the ICMF, disruption and the
cluster-formation efficiency can be better quantified. On the
theoretical front, cosmological simulations will provide a more
detailed picture of galaxy formation and evolution, down to scales
where individual clusters can be followed (e.g., Bournaud \ea 2008;
Prieto \& Gnedin 2008).

\label{lastpage}

\end{document}